# Many-Body Dispersion Interactions in Molecular Crystal Polymorphism

Noa Marom[1], Robert A. DiStasio Jr.[2], Viktor Atalla[3], Sergey Levchenko[3],
James R. Chelikowsky[1], Leslie Leiserowitz[4], and Alexandre Tkatchenko[3]
[1]Center for Computational Materials, Institute for Computational Engineering and Sciences,
The University of Texas at Austin, Austin, TX 78712, USA
[2]Department of Chemistry, Princeton University, Princeton, NJ 08544, USA
[3]Fritz–Haber–Institut der Max–Planck–Gesellschaft, Faradayweg 4-6, 14195, Berlin, Germany
[4]Department of Materials and Interfaces, Weizmann Institute of Science, Rehovoth 76100, Israel
(Dated: September 13, 2012)

Polymorphs in molecular crystals are often very close in energy, yet they may possess markedly different physical and chemical properties. The understanding and prediction of polymorphism is of paramount importance for a variety of applications, including pharmaceuticals, non-linear optics, and hydrogen storage [1, 2]. The crystal structure prediction blind tests conducted by the Cambridge Crystallographic Data Centre have shown steady progress toward reliable structure prediction for molecular crystals [3]. However, several challenges remain, such as molecular salts, hydrates, and flexible molecules with several possible conformations. The ability to rank all the possible crystal structures hinges on a highly accurate description of the relative energetic stability. A first-principles method (i.e., a method based only on the laws of quantum mechanics, which does not require any input from experiment or any ad-hoc assumptions on the nature of the system) that can achieve the required accuracy of 0.1–0.2 kcal/mol per molecule would be an indispensable tool for polymorph prediction. In this Communication, we show that the non-additive many-body dispersion (MBD) energy beyond the standard pairwise approximation is crucial for the correct qualitative and quantitative description of polymorphism in molecular crystals. This is rationalized by the sensitive dependence of the MBD energy on the polymorph geometry and the ensuing dynamic electric fields inside molecular crystals. We use the glycine crystal as a fundamental and stringent benchmark case to demonstrate the accuracy of the MBD method.

Among the available first principles methods, density-functional theory (DFT) is one of the most promising and frequently used approaches to study polymorphism in molecular crystals. However, widely used exchange-correlation functionals (including hybrid functionals) that rely on semi-local correlation fail to capture the contribution of dispersion interation to the stabilization of molecular crystals. These non-bonded interactions are quantum mechanical in nature and physically correspond to the multipole moments induced in response to instantaneous fluctuations in the electron charge density, which is a long-range correlation effect. To incorporate dispersion interactions within DFT, significant progress has been made by utilizing the standard $C_6R^{-6}$ pairwise additive expression for the dispersion energy derived from second-order perturbation theory [4–6]. Indeed, DFT with pairwise dispersion corrections may yield accurate predictions when the energy differences between molecular crystal polymorphs are sufficiently large [7–9]. Most notably, Neumann et al. have achieved the highest success rate in the last two blind tests using such a method [10]. However, the pairwise dispersion energy approaches, even when used in conjunction with state-of-the-art functionals, are still unable to furnish the level of accuracy necessary for describing polymorphism in many relevant molecular crystals [11–14]. In particular, glycine polymorphs are one of the known failures of DFT, as functionals that rely on (semi-)local correlation yield large errors in their lattice parameters and fail to reproduce their relative stability [15]. Recently, an efficient method for describing the many-body dispersion (MBD) energy has been developed [16], building upon the Tkatchenko-Scheffler (TS) method [17]. Within the TS approach the $C_6/R^6$ dispersion term is added in a pairwise fashion to the inter-nuclear energy term. The effective atomic polarizabilities are calculated from first principles, based on the DFT electron density. The MBD method presents a two-fold improvement over the TS approach: (i) the effective polarizability of the system is calculated by solving self-consistently the dipole–dipole electric-field coupling equations; and (ii) the many-body dispersion energy is then calculated to infinite order utilizing the coupled fluctuating dipole model. The inclusion of MBD energy in DFT leads to a significant improvement of binding energies between organic molecules, and for the cohesion of the benzene molecular crystal [16]. The MBD energy, like the TS energy, can be added to any DFT functional, requiring only a once-per-functional adjustment of a range parameter [16, 17].

Glycine (Gly) is the most fundamental of the amino acids, the building blocks of proteins. Beyond its biological and pharmaceutical importance, crystalline glycine is a prototype for hydrogen bonded (H-bonded) networks, an important structural motif in both naturally occurring and artificially engineered molecular crystals. Glycine has three stable polymorphs (See Figure 1): $\alpha$-Gly, $\beta$-Gly, and $\gamma$-Gly. Figure 2 shows the performance of different DFT methods for the prediction of the unit cell volumes of the glycine polymorphs with respect to low temperature experiments. A complete account of the computational details is provided in the supplementary material. As shown in Ref. [15], the local density approximation (LDA) [18] underestimates the unit cell volumes by 7-10%, while the generalized gradient approximation of Perdew, Burke, and Ernzerhof (PBE) [19, 20] overestimates the unit cell volumes by 7-8% [15]. Adding the pairwise TS energy to the PBE functional reduces the error in the unit cell volumes to about 3%, already a significant improvement. PBE+MBD yields further noticeable improvement with an accuracy of 0.3% for the unit cell volumes of $\beta$-Gly and $\gamma$-Gly and 0.8% for $\alpha$-Gly.

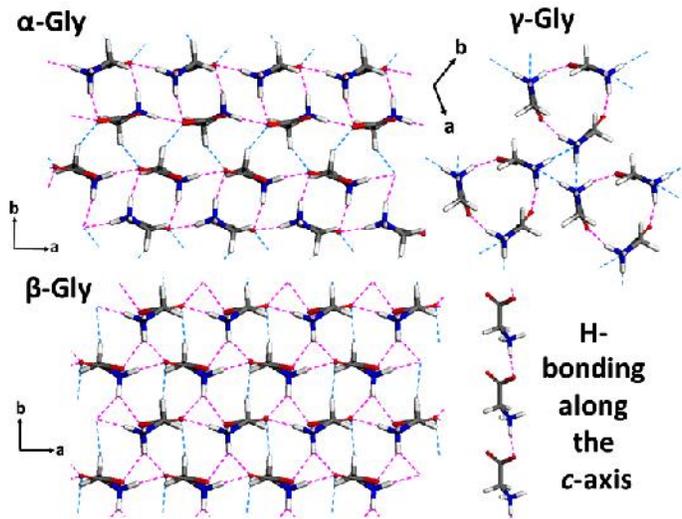

FIG. 1: Structures of the three polymorphs of glycine. H-bonds are indicated by dashed lines. For $\alpha$-Gly (space group $P2_1/n$) and $\beta$-Gly (space group $P2_1$) the strong NH⋯O H-bonds in magenta and the weaker CH⋯O interactions are in cyan. For $\gamma$-Gly (space group $P3_1$) the strong intra-helical H-bonds are in magenta and the weaker inter-helical bonds are in cyan. The translation-related H-bonded chain along the $c$-axis, common to all three polymorphs, is also shown.

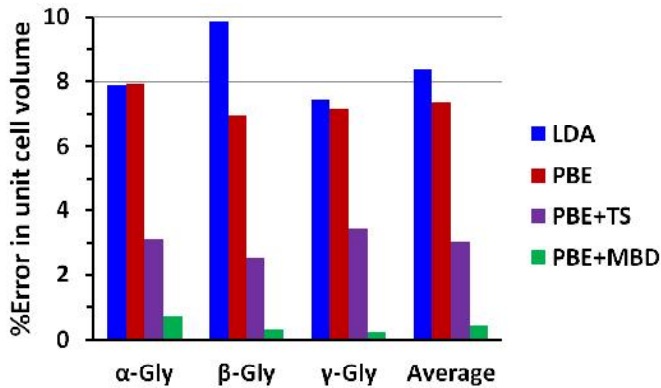

FIG. 2: Percent error in the calculated unit cell volumes of the glycine polymorphs, with respect to low-temperature experiments: $\alpha$-Gly: Refs. [21, 22], $\beta$-Gly: Ref. [23], $\gamma$-Gly: Ref. [24]. LDA and PBE data are from Ref. [15].

Both $\alpha$-Gly and $\beta$-Gly consist of H-bonded sheets of molecular glycine in the $a-c$ plane. The strong H-bonds within the glycine sheets, are described reasonably well by PBE even without including the dispersion energy. This is not the case for the weaker interactions between the glycine sheets, along the $b$ direction. For $\beta$-Gly, in which the glycine sheets are bound by bifurcated NH⋯O bonds, PBE overestimates $b$ by 5%. PBE+TS reduces the overestimation to 1% and PBE+MBD yields excellent agreement with experiment. In $\alpha$-Gly, the glycine sheets form a H-bonded (NH⋯O) bilayer, via the centers of inversion. The three-dimensional (3D) network is then completed by weaker CH⋯O interactions between the bilayers. These interactions determine the direction of the glide (*i.e.*, an $n$-glide as opposed to $a$- or $c$-glide) as well as the inter-bilayer distance along the $b$-axis. The weak interactions along the $b$-direction are reflected in a significant temperature dependence of the $b$ parameter of $\alpha$-Gly [22, 25]. PBE grossly overestimates $b$ by 0.65 Å. PBE+TS significantly reduces the overestimation to 0.16 Å. PBE+MBD does not yield further improvement for the $b$ parameter because the potential energy surface is very flat with the binding energy changing by only 0.01 eV per unit cell for 11.75 Å $< b <$ 12.15 Å.

The most stable $\gamma$-Gly polymorph has the same translation-related H-bonded chain motif as $\alpha$-Gly and $\beta$-Gly along the $c$-axis. However, it is unique in the sense that the H-bonded chains form helices, related by a three-fold screw symmetry, rather than sheets. The helices are held together by lateral NH⋯O H-bonds, forming a 3D network. The inter-helix H-bonds are longer and somewhat weaker than the intra-helix H-bonds. The $c$ parameter is reproduced correctly even by PBE, which successfully captures the strong intra-helix bonds. The $a$ and $b$ parameters are significantly improved by accounting for dispersion interactions. Figure 3 shows the change of the potential energy landscape in the $a$–$b$ plane of $\gamma$-Gly (with $c$ fixed at 5.48 Å), resulting from including the dispersion contributions at different levels of approximation. The TS pairwise dispersion method significantly increases the binding energy and improves the position of the minimum, as compared to standard PBE. However, it is still insufficient for obtaining a highly accurate geometry. Accounting for the MBD interactions correctly captures the weak and complex inter-helix interactions, leading to a slight decrease in the crystal binding energy and yielding a minimum in agreement with experiment. Both of these effects can be explained by accounting for the dynamic electric field inside $\gamma$-Gly, a point that we will elaborate upon below.

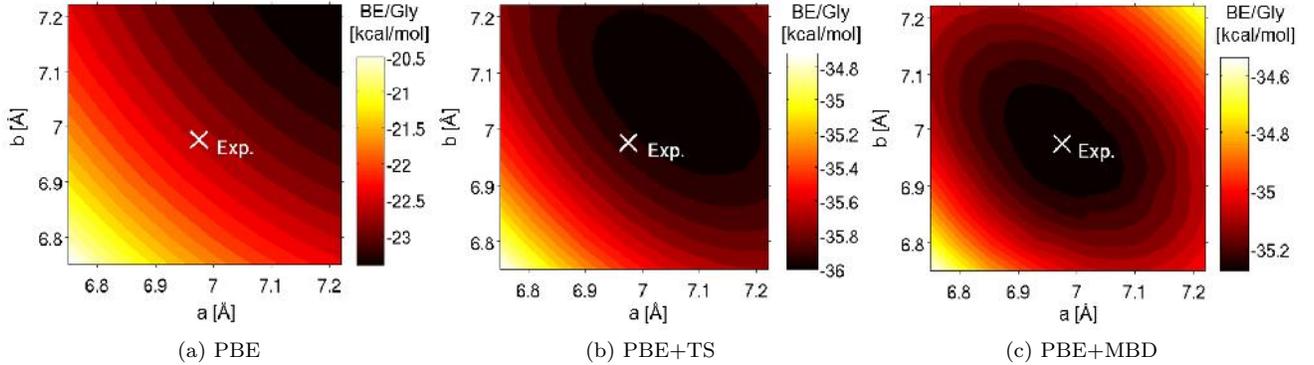

FIG. 3: Potential energy surfaces for the $a$–$b$ plane of $\gamma$-Gly. [26] Experimental lattice parameters are marked by a cross [15].

We now proceed to discuss the relative stability of the glycine polymorphs. Experimentally, it has been determined that $\gamma$-Gly is the most stable polymorph [25], although the energy difference between $\gamma$-Gly and $\alpha$-Gly is very small. It is also well established that $\beta$-Gly is less stable than both $\gamma$-Gly and $\alpha$-Gly [27]. The calculated relative energies are given in Table I and compared to the relative enthalpies from Ref. [25, 28]. LDA predicts the correct order of stability: $\gamma > \alpha > \beta$, but the energy differences between the polymorphs are too large, comparing to experiment. PBE predicts the wrong ordering of the polymorphs with $\alpha$-Gly the most stable and $\gamma$-Gly and $\beta$-Gly nearly degenerate. Both LDA and PBE yield rather large deviations for the lattice constants, thus their relative polymorph energies cannot be trusted. After including the pairwise dispersion energy using the PBE+TS approach, the geometries of the glycine polymorphs improve significantly. However, PBE+TS predicts the wrong order of stability: $\alpha > \beta > \gamma$ and the energy differences between the polymorphs are overestimated. This demostrate yet again that pairwise dispersion corrections fall short when the energy differences between polymorphs are very small. Including the many-body dispersion effects using the PBE+MBD method reproduces the correct order of stability: $\gamma > \alpha > \beta$ and the energy difference between $\alpha$-Gly and $\gamma$-Gly is very close to experiment.

It is known that the zero-point vibrational energy (ZPE) has a non-negligible contribution to the energies of glycine polymorphs [29]. Adding the ZPE to PBE+TS and PBE+MBD stabilizes $\beta$-Gly with respect to the $\alpha$ and $\gamma$ forms, bringing its relative energy closer to experiment. The ZPE contribution to $\alpha$-Gly and $\gamma$-Gly is very similar (see Table I). The PBE-based hybrid functional (PBEh) [30] has been shown to provide an improved description of the stability of hydrogen-bonded ice polymorphs, as compared to PBE [31]. Therefore we also examine the performance of PBEh+MBD for the relative energies of the glycine polymorphs. Indeed, PBEh+MBD further improves the relative stability of the three polymorphs as shown in Table I, yielding very good agreement with experiment. The ZPE correction over PBEh+MBD has the same effect of stabilizing $\beta$-Gly with respect to $\alpha$-Gly and $\gamma$-Gly, as for PBE+MBD. We conclude that the relative stability of the different glycine polymorphs depends critically upon the correct treatment of MBD interactions, with the best results achieved upon the inclusion of ZPE and exact exchange in DFT.

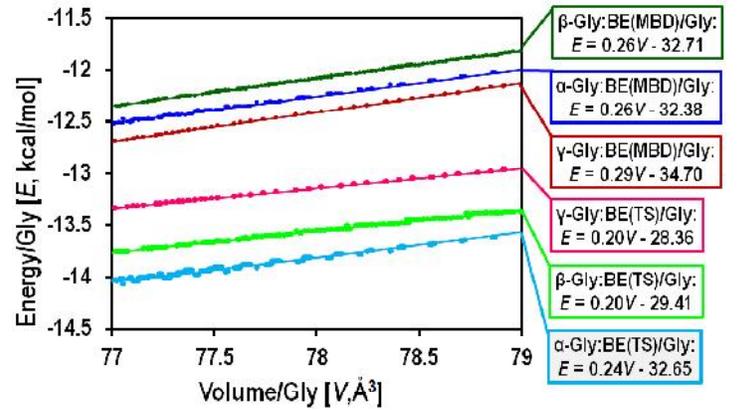

FIG. 4: Dispersion energy contributions using TS and MBD methods to the binding energies of the three polymorphs of glycine as a function of the unit cell volume. For $\alpha$, $\beta$, and $\gamma$ polymorphs, the $a$, $b$, and $c$ unit cell parameters were varied inside the shown unit cell volume range. Both the energy and the volume are normalized per glycine molecule. Parameters for the linear fits are also shown.

Having established the accuracy of the DFT+MBD method for treating the polymorphism of glycine, we analyze the dispersion energy of the three polymorphs in order to explain the stability of $\gamma$-Gly and the role of MBD interactions in its prediction. The plot of the dispersion energy contribution to the crystal binding energy [26] as a function of the unit cell volume in Fig. 4 shows several characteristic features. When using either the TS or MBD methods, the dispersion energy has a linear dependence on the unit cell volume for all the three glycine polymorphs. The MBD energy nicely correlates with the observed experimental stability of the different polymorphs, while the pairwise TS energy fails to do so. In fact, the inclusion of the dipole–dipole screening in the effective polarizability and the many-body effects in the dispersion energy qualitatively changes the energetic ordering and the energy gap between the polymorphs throughout the unit cell volume scan. The linear fits to the dispersion energy offer an additional insight: the $\gamma$-Gly polymorph shows the largest increase of 45% in the slope upon going from the TS to the MBD method. The increased slope explains the observed reduction of 0.1 Å in the $a$ and $b$ lattice parameters with the PBE+MBD method in Fig. 3. The increased stability brought by the MBD energy is clearly sufficient to overcome the intermolecular Pauli repulsion in $\gamma$-Gly and leads to the overall contraction of the unit cell, bringing it into an excellent agreement with low temperature experiments. A significant part of the additional stabilization of the $\gamma$-Gly over the $\alpha$ and $\beta$ polymorphs (see Table I) is attributed to its helical 3D arrangement shown in Fig. 1, which leads to more favorable dipole–dipole screening interactions. As a result of this long-range effect, the polarizability of the glycine molecule in the $\gamma$-Gly crystal is larger by about 1 bohr$^3$ in comparison to the $\alpha$ and $\beta$ polymorphs. This increase in the polarizability corresponds to an increase of about 0.06 in the electronic contribution to the dielectric constant of the glycine crystal. High-precision measurements of the dielectric constant of different glycine polymorphs could be used to confirm our prediction.

TABLE I: Relative energies of glycine polymorphs in kcal/mol per molecule, as obtained with different functionals, compared to measured relative enthalpies from Ref. [25]. ZPE corrected values are given for the dispersion-inclusive methods.

|   | Exp. [25] | LDA [15] | PBE [15] | PBE+TS | PBE+TS+ZPE | PBE+MBD | PBE+MBD+ZPE | PBEh+MBD | PBEh+MBD+ZPE |
|---|---|---|---|---|---|---|---|---|---|
| $\alpha$ | 0.064 | 0.30 | 0 | 0 | 0 | 0.032 | 0 | 0.27 | 0.23 |
| $\beta$ | 0.142 | 0.62 | 0.16 | 0.58 | 0.38 | 0.57 | 0.35 | 0.58 | 0.36 |
| $\gamma$ | 0 | 0 | 0.16 | 0.76 | 0.74 | 0 | 0.011 | 0 | 0 |

To summarize, we found that an accurate description of non-additive many-body dispersion energy with the DFT+MBD method yields the correct structures and relative stability of the $\alpha$, $\beta$, and $\gamma$ glycine polymorphs. The large improvement obtained with the MBD method compared to the simple pairwise model for the dispersion energy is attributed to the sensitive dependence of the MBD energy on the polymorph geometry and the dynamic electric field within molecular crystal. The DFT+MBD method yields an unprecedented accuracy of 0.8% in the description of the structures of glycine polymorphs and of 0.2 kcal/mol in their relative energies. Such accuracy for the energetics is required in order to predict correctly the relative stability of the glycine polymorphs in particular and molecular crystal polymorphs in general.

A.T. acknowledges support from ERC Starting Grant `VDW-CMAT`. Work at UT-Austin was supported under NSF grants DMR-0941645 and OCI-1047997, and DOE grant DE-SC0001878. Computational resources were provided by the Argonne Leadership Computing Facility (ALCF). We thank J. F. Hammond and O. A. von Lilienfeld from ALCF for their interest and help.

# Supplementary Information for: Many-Body Dispersion Interactions in Molecular Crystal Polymorphism

Noa Marom[1], Robert A. DiStasio Jr.[2], Viktor Atalla[3], Sergey Levchenko[3],
James R. Chelikowsky[1], Leslie Leiserowitz[4], and Alexandre Tkatchenko[3]

[1] Center for Computational Materials, Institute for Computational Engineering and Sciences,
The University of Texas at Austin, Austin, TX 78712, USA
[2] Department of Chemistry, Princeton University, Princeton, NJ 08544, USA
[3] Fritz–Haber–Institut der Max–Planck–Gesellschaft, Faradayweg 4-6, 14195, Berlin, Germany
[4] Department of Materials and Interfaces, Weizmann Institute of Science, Rehovoth 76100, Israel

## COMPUTATIONAL DETAILS

All calculations were performed using FHI-AIMS [1, 2], an all-electron numeric atom-centered orbitals (NAO) code. The NAO basis sets are grouped into a minimal basis, containing only basis functions for the core and valence electrons of the free atom, followed by four hierarchically constructed sets of additional basis functions, denoted as *"tier 1-4"*. A detailed description of these basis functions is given in Ref. [1]. Full unit cell relaxations were carried out using the generalized gradient approximation (GGA) of Perdew, Burke, and Ernzerhof (PBE) [3, 4] with the Tkatchenko–Scheffler (TS) dispersion correction [5]. PBE+MBD [6] unit cell optimizations were conducted by scanning the potential energy surface (PES). For each set of lattice parameters, the internal geometry was relaxed with PBE+TS and the MBD energy was calculated subsequently. This approximation is justified because the internal geometries are sufficiently accurate at the PBE+TS level [7, 8]. In both the full unit cell relaxations and the PES scan with internal parameter relaxations, the molecular coordinates were not constrained. Because glycine is a rigid molecule this did not result in significant changes in the molecular geometry. We note that for other systems, comprising larger and more flexible molecular units, accounting for the dispersion interactions in the crystal using the TS correction has led to significant changes in the structure of the molecular unit [7, 8]. Relaxed geometries and the corresponding total energies were determined using a tightly converged *tier 2* basis set. Using a *tier 3* basis set, which corresponds to the basis set limit for DFT calculations, yielded a change of less than 0.02 kcal/mol in the relative energies of the polymorphs with respect to *tier 2*. Additional single-point energy evaluations, using the one-parameter PBE-based hybrid functional, PBEh [9], were performed at the PBE+MBD optimized geometries.

TABLE I: Cell parameters in Å of the glycine polymorphs, as obtained with different methods, compared to low temperature experiments. (a) $\alpha$-Gly: Refs. [10, 11] , $\beta$-Gly: Ref. [12] , $\gamma$-Gly: Ref. [13]. (b) Ref. [14]. The local density approximation (LDA) [15] underestimates the unit cell volumes by 7-10%, while PBE overestimates the unit cell volumes by 7-8% [14]. Adding the pairwise TS energy to the PBE functional reduces the error in the unit cell volumes to about 3%, a significant improvement over LDA and PBE. PBE+MBD further improves upon PBE+TS, yielding geometries in excellent agreement with experiment. For $\beta$-glycine and $\gamma$-glycine the unit cell volume is accurate to 0.3%. The unit cell volume of $\alpha$-glycine is accurate to 0.8%.

|  | Exp.[a] | LDA[b] | PBE[b] | PBE+TS | PBE+MBD |
|---|---|---|---|---|---|
| $\alpha$-glycine | | | | | |
| a | 5.09 | 4.96 | 5.15 | 5.14 | 5.08 |
| b | 11.77 | 11.34 | 12.42 | 11.93 | 11.92 |
| c | 5.46 | 5.36 | 5.44 | 5.46 | 5.44 |
| volume | 303.2 | 279.3 | 327.3 | 312.6 | 305.4 |
| $\beta$-glycine | | | | | |
| a | 5.08 | 4.93 | 5.11 | 5.10 | 5.06 |
| b | 6.18 | 5.82 | 6.47 | 6.24 | 6.18 |
| c | 5.39 | 5.33 | 5.39 | 5.40 | 5.39 |
| volume | 155.2 | 139.9 | 166.0 | 159.1 | 154.7 |
| $\gamma$-glycine | | | | | |
| a | 6.98 | 6.76 | 7.21 | 7.08 | 6.97 |
| b | 6.98 | 6.76 | 7.21 | 7.09 | 6.97 |
| c | 5.47 | 5.39 | 5.48 | 5.48 | 5.47 |
| volume | 230.6 | 213.4 | 247.1 | 238.5 | 230.1 |

## CELL PARAMETERS